\documentclass[10pt]{amsart}
\usepackage{latexsym,amssymb,amsmath,amsthm,amscd,
epic,
graphics,amsxtra
}
\renewcommand{\div}{div}
\newcommand{\Lie}{Lie}

\newcommand{\mx}[1]{\mbox{{#1}}}

\DeclareMathOperator{\End}{End}
\DeclareMathOperator{\high}{high}
\DeclareMathOperator{\Hom}{Hom}

\DeclareMathOperator{\Ker}{Ker}
\DeclareMathOperator{\low}{low}

\DeclareMathOperator{\tr}{tr}

\newcommand{\1}{\boldsymbol{1}}

\def\CC{{\mathbb{C}}}

\def\ZZ{{\mathbb{Z}}}

\def\Bw{{\mathcal B}}

\def\Hw{{\mathcal H}}

\newcommand{\fg}{\mbox{{\tt g}}}

\newcommand{\ad}{\mathop{\rm ad \, }}

\renewcommand{\tilde}{\widetilde}
\renewcommand{\theequation}%
  {\arabic{section}.\arabic{equation}}
\makeatletter
\renewcommand\section%
 {\@startsection {section}{1}{\z@}%
  {-3.5ex \@plus -1ex \@minus -.2ex}%
    {2.3ex \@plus.2ex}%
       {\normalfont\large\bfseries}}
\@addtoreset{equation}{section}
\makeatother

\newtheorem{proposition}{Proposition}[section]

\newtheorem{Proposition}{Proposition}[section]
\newcommand{\bPr}{\begin{Proposition}}
\newcommand{\ePr}{\end{Proposition}}

\newtheorem{Theorem}[Proposition]{Theorem}
\newcommand{\bTh}{\begin{Theorem}}
\newcommand{\eTh}{\end{Theorem}}

\newtheorem{Lemma}[Proposition]{Lemma}
\newcommand{\bLe}{\begin{Lemma}}
\newcommand{\eLe}{\end{Lemma}}

\newtheorem{Definition}[Proposition]{Definition}
\newcommand{\bDe}{\begin{Definition}}
\newcommand{\eDe}{\end{Definition}}

\newtheorem{Corollary}[Proposition]{Corollary}
\newcommand{\bCo}{\begin{Corollary}}
\newcommand{\eCo}{\end{Corollary}}

\newtheorem{Conjecture}[Proposition]{Conjecture}
\newcommand{\bCj}{\begin{Conjecture}}
\newcommand{\eCj}{\end{Conjecture}}

\theoremstyle{remark}
\newtheorem{remark}[Proposition]{Remark}

\newcommand{\bRe}{\begin{remark}}
\newcommand{\eRe}{\end{remark}}

\newcommand{\bEq}{\begin{equation}}
\newcommand{\eEq}{\end{equation}}

\newcommand{\bEa}{\begin{eqnarray}}
\newcommand{\eEa}{\end{eqnarray}}

\newcommand{\bEaz}{\begin{eqnarray*}}
\newcommand{\eEaz}{\end{eqnarray*}}

\newcommand{\bAr}{\begin{array}}
\newcommand{\eAr}{\end{array}}

\newcommand{\bN}{\begin{enumerate}}
\newcommand{\eN}{\end{enumerate}}

\newcommand{\bD}{\begin{description}}
\newcommand{\eD}{\end{description}}

\begin{document}


\title[Complexes of modules over exceptional Lie superalgebras $E
(3,8)$ and $E (5,10)$.]
{Complexes of modules over exceptional Lie superalgebras $E
(3,8)$ and $E (5,10)$.}

\author{Victor G. Kac${}^*$ and Alexei Rudakov}
\thanks{${}^*$~Supported in part by NSF grant
    DMS-9970007. \\
}

\begin{abstract}
In this paper complexes of generalized Verma modules over the
infinite-dimensional exceptional Lie superalgebras $E (3,8)$ and
$E(5,10)$ are constructed and studied.

\end{abstract}
\maketitle

\section{Introduction.}

In our papers \cite{KR1}--\cite{KR3} we constructed all
degenerate irreducible modules over the exceptional Lie
superalgebra $E(3,6)$.  In the present paper we apply the same
method to the exceptional Lie superalgebras $E(3,8)$ and
$E(5,10)$.

The Lie superalgebra $E(3,8)$ is strikingly similar to $E(3,6)$.
In particular, as in the case of $E(3,6)$, the maximal compact
subgroup of the group of automorphisms of $E (3,8)$ is isomorphic
to the group of symmetries of the Standard Model.  However, as
the computer calculations by Joris van Jeugt show, the
fundamental particle contents in the $E (3,8)$ case is completely
different from that in the $E(3,6)$ case \cite{KR2}.  All the
nice features of the latter case, like the CPT symmetry,
completely disappear in the former case.  We believe that the
main reason behind this is that, unlike $E(3,6)$, $E(3,8)$ cannot
be embedded in $E(5,10)$, which, we believe, is the algebra of
symmetries of the $SU_5$ Grand Unified Model (the maximal compact
subgroup of the automorphism group of $E(5,10)$ is $SU_5$).

However, similarity with $E(3,6)$ allows us to apply to $E(3,8)$
all the arguments from \cite{KR2} almost verbatim, and Figure~1
 of the present paper that depicts all degenerate $E
 (3,8)$-modules is almost the same as Figure~3 from \cite{KR2}
 for $E(3,6)$.

The picture in the $E(5,10)$ case is quite different (see
Figure~2).  We believe that it depicts all degenerate irreducible
$E(5,10)$-modules, but we still do not have a proof.

\section{Morphisms between generalized Verma modules.}
\label{sec:1}
Let $L = \oplus_{j \in \ZZ} \fg_j$ be a $\ZZ$-graded Lie
superalgebra by finite-dimensional vector spaces.  Let
\begin{displaymath}
  L_- = \oplus_{j<0}\,\, \fg_j , \,\,\, L_+ = \oplus_{j>0} \,\,\fg_j ,
  \,\,\, L_0 = \fg_0 + L_+ \, .
\end{displaymath}
Given a $\fg_{{0}}$-module $V$, we extend it to a $L_{{0}}$-module by
letting $L_+$ act trivially, and define the induced $L$-module
\begin{displaymath}
  M(V) = U(L) \otimes_{U{(L_0)}} V \, .
\end{displaymath}

If $V$ is a finite-dimensional irreducible  $\fg$-module, the
$L$-module $M(V)$ is called a \emph{generalized Verma module}
(associated to $V$), and it is called \emph{degenerate} if it is
not irreducible.

Let $A$ and $B$ be two $\fg_0$-modules and let $\Hom (A,B)$ be
the $\fg_0$-module of linear maps from $A$ to $B$.  The following
proposition will be extensively used to construct morphisms
between the $L$-modules $M(A)$ and $M(B)$.

\begin{proposition}
  \label{prop:1.1}
  Let $\Phi \in M(\Hom (A,B))$ be such that
  \begin{equation}
    \label{eq:1.1}
    v \cdot \Phi =0 \hbox{ for all } v \in L_0 \, .
  \end{equation}
Then we can construct a well-defined morphism of $L$-modules
\begin{displaymath}
  \varphi : M(A) \to M(B)
\end{displaymath}
by the rule $\varphi (u \otimes a) = u\,\Phi (a)$.  Explicitly,
write $\Phi = \sum_m u_m \otimes \ell_m$, where $u_m \in U(L)$,
$\ell_m \in \Hom (A,B)$.  Then
\begin{equation}
  \label{eq:1.2}
 \varphi (u \otimes a) = \sum_m (uu_m ) \otimes \ell_m (a) \, .
\end{equation}

\end{proposition}

\begin{proof}
  We have to prove that for $v \in U (L_0)$,
  \begin{displaymath}
   \varphi (uv \otimes a) = \varphi (u \otimes va) \, ,
  \end{displaymath}
in order to conclude that $\varphi$ is well-defined.  Notice that
condition (\ref{eq:1.1}) means
\begin{equation}
  \label{eq:1.3}
  \sum_m [v, u_m] \otimes \ell_m + \sum_m u_m \otimes v \ell_m =0
  \, .
\end{equation}
Therefore we have:
\begin{eqnarray*}
  \varphi (uv \otimes a)
  &=& \sum_m uvu_m \otimes \ell_m (a) \\
  &=& \sum_m u [v,u_m] \otimes \ell_m (a)
      + \sum_m uu_m v \otimes \ell_m (a)\\
  &=& \sum_m u [v,u_m] \otimes \ell_m (a)
      + \sum_m uu_m \otimes v (\ell_m (a))\\
  &=& \sum_m u [v,u_m] \otimes \ell_m (a)
      + \sum_m uu_m \otimes (v \ell_m) (a)
      + \sum_m uu_m \otimes \ell_m (va)\\
\hbox{(by (\ref{eq:1.3}) and (\ref{eq:1.2}))}
  &=& \sum uu_m \otimes \ell_m (va)
     \,\, =\,\, \varphi (u \otimes va) \, .
\end{eqnarray*}
The fact that $\varphi$ defines a morphism of $L$-modules is
immediate from the definition.
\end{proof}

\begin{remark}
  \label{rem:1.2}
If $L_0$ is generated by $\fg_0$ and a subset of $T \subset L_+$,
then condition (\ref{eq:1.1}) is equivalent to
\begin{subequations}
\begin{eqnarray}
  \label{eq:1.4a}
\fg_0 \cdot \Phi &=& 0 \, \\
\label{eq:1.4b}
   a \cdot \Phi &=& 0 \,\,\hbox{ for all }\, a \in T \, .
\end{eqnarray}
\end{subequations}
Condition (\ref{eq:1.4a}) usually gives a hint to a possible
shape of $\Phi$ and is checked by general invariant-theoretical
considerations.  After that (\ref{eq:1.4b}) is
usually checked by a direct calculation.
\end{remark}

\begin{remark}
  \label{rem:1.3}
We can view $M(V)$ also as the induced $(L_- \oplus \fg_0)$-module:
$U(L_- \oplus \fg_0) \otimes_{U(\fg_0)}V$.  Then condition
(\ref{eq:1.4a}) on $\Phi = \sum_m u_m \otimes \ell_m$, where $u_m
\in U(L_- \oplus \fg_0)$, $\ell_m \in \Hom (A,B)$, suffices in order
(\ref{eq:1.2}) to give a well-defined morphism of
$(L_- \oplus \fg_0)$-modules.  One can also replace $\fg_0$ by any of its
subalgebras.

\end{remark}



\section{Lie superalgebras $E(3,6)$, $E (3,8)$ and $E(5,10)$.}
\label{sec:2}

Recall some standard notation:
\begin{displaymath}
  W_n = \{ \sum^n_{j=1} P_i (x) \partial_i\,\,
| \,\, P_i \in \CC [[x_{\bar{1}} , \ldots ,x_n ]], \,\,
  \partial_i \equiv \partial / \partial x_i \}
\end{displaymath}
denotes the Lie algebra of formal vector fields in $n$
indeterminates:
\begin{displaymath}
  S_n = \{ D = \sum P_i \partial_i  \, \, |\,\,
  \div D \equiv \sum_i \partial_i P_i =0 \}
\end{displaymath}
denotes the Lie
  subalgebra of divergenceless formal vector fields; $\Omega^k (n)$
  denotes the associative algebra of formal differential forms of
  degree $k$ in $n$ indeterminates, $\Omega^k_{c\ell} (n)$ denoted
  the subspace of closed forms.

The Lie algebra $W_n$ acts on $\Omega^k (n)$ via Lie derivative
$D \to L_D$.  Given $\lambda \in \CC$ one can twist this action:
\begin{displaymath}
  D \,\omega = L_D \,\omega + \lambda (\div D)\, \omega \, .
\end{displaymath}
The $W_n$-module thus obtained is denoted by $\Omega^k
(n)^{\lambda}$.  Recall the following obvious
isomorphism of $W_n$-modules
\begin{equation}
  \label{eq:2.1}
  \Omega^0 (n) \simeq \Omega^n (n)^{-1} \, ,
\end{equation}
and the following slightly less obvious  isomorphism of
$W_n$-modules
\begin{equation}
  \label{eq:2.2}
  W_n \simeq \Omega^{n-1} (n)^{-1} \, .
\end{equation}
The latter is obtained by mapping a vector field $D \in W_n$ to
the {\footnotesize{$(n-1)$}}-form
$\iota_D (dx_{{1}} \wedge \ldots \wedge dx_n)$.  Note that
(\ref{eq:2.2}) induces an isomorphism of $S_n$-modules:
\begin{equation}
  \label{eq:2.3}
  S_n \simeq \Omega^{n-1}_{c\ell} (n) \, .
\end{equation}

Recall that the Lie superalgebra $E(5,10) = E(5,10)_{\bar{0}} + E
(5,10)_{\bar{1}}$ is constructed as follows \cite{K}, \cite{CK}:
\[
E(5,10)_{\bar{0}} = S_5, \quad E(5,10)_{\bar{1}} = \Omega^2_{c\ell}(5),
\]
$E(5,10)_{\bar{0}}$ acts on $E(5,10)_{\bar{1}}$ via the Lie derivative,
 and
$\,[\omega_2 , \omega'_2] = \omega_2 \wedge \omega'_2 \in
\Omega^4_{c\ell} (5) =S_5\,$ (see (\ref{eq:2.3}) ) $\,$ for
$\,\,\omega_2,\omega'_2 \in E(S,10)_{\bar{1}}$.

Next, recall the construction of the Lie superalgebras
$E^{\flat}:=E(3,6)$ and $E^{\sharp}:=E(3,8)$ \cite{CK}:
\begin{eqnarray*}
  E^{\flat}_{\bar{0}} &=& E^{\sharp}_{\bar{0}}
 = W_3 + s \ell_2 (\Omega^0(3) )\,\,
  \hbox{ (the natural semidirect sum)}\, ;\\
  E^{\flat}_{\bar{1}} &=& \Omega^1 (3)^{-1/2} \otimes \CC^2 ,\,\,\,\,
      E^{\sharp}_{\bar{1}} = (\Omega^0(3)^{-1/2} \otimes \CC^2)
      + (\Omega^2 (3)^{-1/2} \otimes \CC^2) \, .
\end{eqnarray*}
The action of the even on the odd parts is defined via the Lie
derivative and the multiplication of a function and a
differential form.  The bracket of two odd elements is defined by
using the identifications (\ref{eq:2.1}) and (\ref{eq:2.2}) as
follows.  For $\omega_i, \omega'_i \in \Omega^i(3) $ and $v,v'
\in \CC^2$ one defines the following bracket of two elements from
$E^{\flat}_{\bar{1}}$:
\begin{equation}
  \label{eq:2.4}
  [\omega_{{1}} \otimes v , \omega'_{{1}} \otimes v']
    = - (\omega_{{1}} \wedge \omega'_{{1}}) \otimes (v \wedge v')
    - (d \omega_{{1}} \wedge \omega'_{{1}}
                + \omega_{{1}} \wedge d\omega'_{{1}})
    \otimes (v \cdot v') \, ,
\end{equation}
and the following bracket of two elements from $E^{\sharp}_{\bar{1}}$:
\begin{eqnarray}
  \label{eq:2.5}
  [\omega_2 \otimes v , \omega'_2 \otimes v'] &=& 0 ,\,\,\,\,
  [\omega_0 \otimes v,\omega'_0 \otimes v'] =
  -(d\omega_0 \wedge d\omega'_0) \otimes (v \wedge v'),\\
\nonumber
  [\omega_0 \otimes v , \omega_2 \otimes v'] &=&
  - (\omega_0\wedge \omega_2) \otimes (v \wedge v')
  - (d\omega_0 \wedge \omega_2 -\omega_0 \wedge\omega_2)
  \otimes (v \cdot v')) \, .
\end{eqnarray}

Recall also an embedding of $E^{\flat}$ in $E (5,10)$ \cite{CK},
\cite{KR2}.  For that let $z_+ = x_4 $, $z_- = x_5$, $\partial_+ =
\partial_4$, $\partial_- =\partial_5$, and let $\epsilon^+,
\epsilon^-$ denote the standard basis of $\CC^2$.  Then
$E^{\flat}_0$ is embedded in $E (5,10)_0 =S_5$ by $(D \in W_3 ,
a,b,c \in \Omega^0 (3))$:
\begin{equation}
  \label{eq:2.6}
  D \mapsto D -\tfrac{1}{2} (\div D) (z_+ \partial_+ + z_-
  \partial_-), \,\,
  \left(
    \begin{array}{lr}
       a& b \\ c & -a
    \end{array}
  \right)
 \mapsto a (z_+ \partial_+ -z_-
  \partial_-)
  + bz_+ \partial_- + cz_- \partial_+ \, ,
\end{equation}
and $E^{\flat}_{\bar{1}}$ is embedded in $E(5,10)_{\bar{1}} =
\Omega^2_{c\ell}(5)$ by $(f \in \Omega^0(3))$:
\begin{equation}
  \label{eq:2.7}
  f \, dx_i \otimes \epsilon^{\pm} \mapsto z_{\pm} \,
  dx_i \wedge \, df + f \, dx_i \wedge \, dz_{\pm} \, .
\end{equation}

Introduce the following subalgebras $S^{\flat} \subset E^{\flat}$
and $S^{\sharp} \subset E^{\sharp}$:
\begin{eqnarray*}
  S^{\flat}_{\bar{0}} &=& S^{\sharp}_{\bar{0}}
       \,\,\,= \,\,\, W_3 + \CC \otimes s\ell_2 (\CC)
  \, , \\
  S^{\flat}_{\bar{1}} &=& \Omega^1_{c\ell} (3)^{-1/2} \otimes \CC^2 \, ,\,
     \,\, S^{\sharp}_{\bar{1}} = \Omega^0 (3)^{-1/2} \otimes \CC^2 \, .
\end{eqnarray*}

\begin{Proposition}
  \label{prop:2.1}
  The map $S^{\sharp} \to S^{\flat}$, which is identical on
  $S^{\sharp}_{\bar{0}}$ and sends $f \otimes v \in S^{\sharp}_{\bar{1}}$ to $df
  \otimes v \in S^{\flat}_{\bar{1}}$ is a surjective homomorphism of Lie
  superalgebras with $2$-dimensional central kernel $\CC \otimes
  \CC^2 \subset S^{\sharp}_{\bar{1}}$.
\end{Proposition}

\begin{proof} It is straightforward using (\ref{eq:2.4}) and (\ref{eq:2.5}).

\end{proof}

This proposition is probably the main reason for a remarkable
similarity between representation theories of $E^{\sharp}$ and
$E^{\flat}$.  We shall stress this similarity in our notation and
develop representation theory of $E^{\sharp}$ along the same
lines as that of $E^{\flat}$ done in \cite{KR1}, \cite{KR2}.
Sometimes we shall drop the superscript $\flat$ or $\sharp$ when
the situation is the same.

Recall that $E^{\flat}$ carries a unique irreducible consistent
$\ZZ$-gradation. It has depth $2$, and it is defined by:
\begin{equation}
  \label{eq:2.8}
  \deg x_i = -\deg \partial_i =2 \, , \,\,
  \deg \partial_i =-1 \, , \,\, \deg \epsilon^{\pm} =0 \, , \,\,
  \deg s\ell_2 (\CC) =0 \, .
\end{equation}
The ``non-positive'' part of this $\ZZ$-gradation is as follows:
\begin{eqnarray*}
  \fg_{-2} &=& \langle \partial_i \, | i = 1,2,3 \rangle \, , \,\,
  \fg_{-1}\,\, =\,\, \langle d^a_i :=\epsilon^a \otimes dx_i
     = dx_i \wedge dz_a \,| i=1,2,3,a=+,- \rangle \, ,\\
     \fg_0 &=& s\ell_3 (\CC) \oplus s\ell_2 (\CC) \oplus CY \, ,
\end{eqnarray*}
where
\begin{eqnarray}
  \label{eq:2.9}
  s \ell_3 (\CC) &=& \langle h_1 = x_1 \partial_1 -x_2 \partial_2 ,\,
       h_2 =x_2 \partial_2 -x_3 \partial_3 ,\,
       e_1 =x_1 \partial_2,\\
\nonumber
  &&\,\,\,e_2 = x_2 \partial_3 ,\, e_{12} = x_{{1}} \partial_3 ,\,
        f_{{1}} =x_2 \partial_{{1}} ,\, f_2=x_3\partial_2 ,\,
        f_{12} =x_3\partial_{{1}} \rangle \, , \\
  \label{eq:2.10}
   s\ell_2 (\CC) &=& \langle h_3 = z_+ \partial_+ -z_- \partial_- ,\,\,
        e_3 = z_+ \partial_- , \,f_3 =z_-\partial_+ \rangle \, ,\\
  \label{eq:2.11}
  Y &=&\tfrac{2}{3} \sum x_i\partial_i
        - (z_+\partial_+ + z_- \partial_-) \, .
\end{eqnarray}
The eigenspace decomposition of $\ad (3Y)$ coincides with the
consistent $\ZZ$-grading of $E^{\flat}$.
We fix the Cartan subalgebra
$\Hw = \langle h_{{1}},h_2,h_3,Y \rangle$
and the Borel subalgebra
 $\Bw=\Hw + \langle e_i ,\,
\hbox{\footnotesize{ 
$(i=1,2,3)$}},\,\,
e_{12} \rangle$ of $\fg_0$.  Then $f_0 := d^+_1$ is the highest
weight vector of the (irreducible) $\fg_0$-module
$\fg_{-1}$, the vectors
\begin{displaymath}
  e'_0 := x_3 d^-_3 \,\, \hbox{  and  } \,\, e^{\flat}_0 :=x_3 d^-_2 -
     x_2 d^-_3 + 2z_- dx_2 \wedge dx_3
\end{displaymath}
are all the lowest weight vectors of the $\fg_0$-module $\fg_1$, and
one has:
\begin{eqnarray}
  \label{eq:2.12}
  [e'_0 , f_0] &=& f_2 \, , \\
  \label{eq:2.13}
  [e^{\flat}_0 ,f_0] &=& \tfrac{2}{3} h_1 + \tfrac{1}{3} h_2 -h_3-Y
       =:h^{\flat}_0
  \, , \\
\noalign{\nonumber\hbox{so that } }
h^{\flat}_0 &=& -x_2 \partial_2 -x_3 \partial_3 +2z_-\partial_- \, .
%
\end{eqnarray}
The following relations are also important to keep in mind:
\begin{eqnarray}
  \label{eq:2.15}
  [\,e'_0 , d^+_1] &\!\!= f_2 ,& \,
      [e'_0 ,d^+_2]  =-f_{12}, \,\,\,
      [e'_0 ,d^+_3] =0 , \,\,\,
  [e'_0 , d^-_i] = 0 \, , \\
  \label{eq:2.16}
  [\,d^{\pm}_i , d^{\pm}_j] &\!\!=0\, ,&\,\,\quad
  [d^+_i ,d^-_j] + [d^+_j , d^-_i] = 0 \, .
\end{eqnarray}
Recall that $\fg_0$ along with the elements $f_0$, $e^{\flat}_0$, $e'_0$
generate the Lie superalgebra $E^{\flat}$ \cite{CK}.

The Lie superalgebra $E^{\sharp}$ carries a unique consistent
irreducible
$\ZZ$-gradation of depth $3$:
\begin{displaymath}
E^{\sharp} = \oplus_{j \geq -3}\,\, \fg_j \, .
\end{displaymath}
It is defined by:
\begin{equation}
  \label{eq:2.17}
  \deg x_i = -\deg \partial_i = \deg \, dx_i =2, \,\,\,
  \deg \epsilon^{\pm} =-3, \,\,\, \deg s\ell_2 (\CC)=0 \, .
\end{equation}
In view of Proposition~\ref{prop:2.1} and the above
$E^{\flat}$-notation, we introduce the following
$E^{\sharp}$-notation:
\begin{eqnarray}
  \label{eq:2.18}
  &d^{\pm}&\!\! := 1  \otimes \epsilon^{\pm}  , \,\,
  d^{\pm}_i := x_i \otimes \epsilon^{\pm}  , \,\,
  e'_0 := \tfrac{1}{2} x^2_3 \otimes \epsilon^{\pm}  , \,\,
  f_0 := d^+_{{1}} \, , \\
\nonumber
&e^{\sharp}_0 &\!\! :=  -(dx_2 \wedge dx_3) \otimes \epsilon^-, \, \,\,
   h^{\sharp}_0 := \tfrac{2}{3} h_1  +  \tfrac{1}{3} h_2
      - \tfrac{1}{2} h_3 + \tfrac{1}{2} Y \, .
\end{eqnarray}
If, in analogy with $E^{\flat}$, we denote $\epsilon^a = dz_a$,
then $\fg_{-3}= \langle dz_+ , dz_- \rangle$, $\fg_{-2}$,
$\fg_{-1}$ and $\fg_0$ are the same as for $E^{\flat}$ (except
that now $[\fg_{-1}, \fg_{-2}] \neq 0$), and the relations
(\ref{eq:2.12}), (\ref{eq:2.13}),
(\ref{eq:2.15}), (\ref{eq:2.16})  still hold, but the formula
for $h_0$ is different:
\begin{equation}
  \label{eq:2.19}
  h^{\sharp}_0 = \tfrac{2}{3} h_1 + \tfrac{1}{3} h_2 -
     \tfrac{1}{2}h_3 + \tfrac{1}{2}Y =
     x_1 \partial_1 -z_+\partial_+ \, .
\end{equation}

As in the $E^{\flat}$ case, the elements $e_i$, $f_i$, $h_i$ for
$i=0,1,2,3$ along with $e'_0$ generate $E^{\sharp}$, the elements
$e^{\sharp}_0$ and $e'_0$ are all lowest weight vectors of the
$\fg_0$-module $\fg_1$, and $\fg_0$ along with $e'_0$ and
$e^{\sharp}_0$ generate the subalgebra
$\oplus_{j \geq 0}\,\fg_j$ \cite{CK}.
Thus, by Remark~\ref{rem:1.2}, condition~(\ref{eq:1.4b}) is
equivalent to
\begin{equation}
  \label{eq:2.20}
  e'_0 \cdot \Phi =0 \quad \hbox{and} \quad
  e^{\sharp}_0 \cdot \Phi =0 \, .
\end{equation}

\section{Complexes of degenerate Verma modules over $E(3,6)$.}
\label{sec:3}

Let $W$ be a finite-dimensional symplectic vector space and let
$H$ be the corresponding Heisenberg algebra
\begin{displaymath}
  H=T (W) /(v \cdot w - w \cdot v -(v,w) \cdot 1)
\end{displaymath}
where $T(W)$ denotes the tensor algebra over $W$ and $(~~,~~)$ is the
non-degenerate symplectic form on $W$.  Given two transversal
Lagrangian subspaces $L,L' \subset W$, we have a canonical
isomorphism of symplectic spaces:  $W = L+L' \simeq L \oplus
L^*$, and we can canonically identify the symmetric algebra $S
(L)$ with the factor of $H$ by the left ideal generated by $L'$:
\begin{equation}
\label{eq:3.1}
  V_L := H/(L') \simeq S(L) \1_L \, , \,\, \hbox{ where }
  \1_L = 1+ L' \, .
\end{equation}
One thus acquires an $H$-module structure on $S(L)$.

We construct a symplectic space $W$ by taking
$x_{{1}},x_2,x_3,z_+,z_-,\partial_{{1}},\partial_2,\partial_3,
\partial_+,\partial_-$
as a basis with the first half being dual to the second half:
\begin{equation}
  \label{eq:3.2}
  (\partial_i , x_j) =- (x_j , \partial_i) = \delta_{ij} \, , \,\,
  (\partial_a,z_b) =-(z_b,\partial_a) =\delta_{a,b} \, , \,\,
  \hbox{ all other pairings zero.}
\end{equation}

In general the decomposition $W=L+L' \simeq L \oplus L^*$
provides the canonical maps:
\begin{displaymath}
  \End (L) \overset{\sim}{\to} L \otimes L^*  \overset{\sim}{\to}
  L \cdot L' \hookrightarrow H \, ,
 \end{displaymath}
which induce a Lie
  algebra homomorphism:  $ g\ell (L) \to H_{\Lie} $,
where the Lie algebra structure is defined by the usual commutator.

We consider the following subspaces of $W$:
\begin{equation}
  \label{eq:3.3}
\begin{array}{ll}
  L_A = L'_D = \langle x_i , z_a \rangle \, ,& \,\,
  L_B = L'_C = \langle x_i , \partial_a \rangle \, , \,\,\\
  L_C = L'_B = \langle \partial_c,z_a \rangle \, , & \,\,
  L_D = L'_A  = \langle \partial_i ,\partial_a \rangle \, , \,
\end{array}
\end{equation}
where $ i=1,2,3 \, , \,\, a={\hbox{\footnotesize{$+,-$}}}$.
Note that these are the only $\fg_0$-invariant Lagrangian subspaces of $W$.

As formulae (\ref{eq:2.9})--(\ref{eq:2.11}) determine the
  inclusion $\fg_0 \hookrightarrow g\ell (V_X)$, where $X=A$, $B$, $C$ or
$ D$, we get a Lie algebra monomorphism:
 \begin{equation}
   \label{eq:3.4}
   \fg_0 \hookrightarrow \fg \ell (V_X) \hookrightarrow H_{\Lie} \, .
 \end{equation}
Thus we get a $\fg_0$-action on $V_X$.  Let us notice that by
(\ref{eq:3.4})
\begin{equation}
  \label{eq:3.5}
  Y \to Y^{\flat} = \tfrac{2}{3} (\sum_i x_i \partial_i)-
  (\sum_a z_a \partial_a)
\end{equation}
and
\begin{displaymath}
Y^{\flat} \,\1_A =0  \, , \,\,\, Y^{\flat} \,\1_B = 2 \,\1_B \, ,\,\,\,
Y^{\flat} \,\1_C = -2\,\1_C\, , \,\,\, Y^{\flat} \,\1_D =0 \, ,
\end{displaymath}
as it should be for
$E^{\flat}=E(3,6)$ (see \cite{KR2}).

In the $E(3,8)$ case we modify the $\fg_0$-action on $V_X$
leaving it the same for $s\ell_3 (\CC) \oplus s\ell_2 (\CC)
\subset \fg_0$, but letting
\begin{equation}
  \label{eq:3.6}
  Y \mapsto Y^{\sharp} = -\tfrac{4}{3}(\sum_i x_i \partial_i)
    + (\sum_az_a\partial_a) =Y^{\flat} +2T \, , \,\,
    \hbox{ where } T= -\sum_i x_i\partial_i +\sum_az_a\partial_a \, .
\end{equation}
We have got:
\begin{eqnarray*}
Y \, x^p_i z^r_a  \,\1_A
 &=& \left( -\tfrac{4}{3} p+r \right) x^p_i z^r_a  \,\1_A \, ,\\[1ex]
Y  \,x^p_i \partial^r_a \,\1_B
 &=& \left( \tfrac{4}{3}p-r-2 \right) x^p_i \partial^r_a
 \,\1_B \, , \\[1ex]
Y  \,\partial^q_i  z^r_a \, \1_C
 &=& \left( \tfrac{4}{3}q+r+4 \right)\partial^q_i z^r_a  \,\1_C
 \, , \\[1ex]
Y \, \partial^q_i \partial^r_a  \,\1_D
 &=& \left( \tfrac{4}{3}q-r+2 \right) \partial^q_i \partial^r_a
 \,\1_D \, .
\end{eqnarray*}

Let $F(p,q;r;y)$ denote the finite-dimensional irreducible
$\fg_0$-module with highest weight $(p,q;r;y)$, where $p,q,r \in
\ZZ_+$, $y \in \CC$, and let $M(p,q;r;y) =M (F(p,q;r;y))$ be the
corresponding generalized Verma module over $E(3,8)$.  This
module has a unique irreducible quotient denoted by $I
(p,q;r;y)$.  The latter module is called degenerate if the former
is.

We announce below a classification of all degenerate irreducible
$E(3,8)$-modules.

\begin{Theorem}
  \label{th:1}
All irreducible degenerate $E(3,8)$-modules $I(p,q;r,y)$ are as follows\\
$\hbox{$(p,q,r \in \ZZ_+):$}$
\begin{eqnarray*}
  \begin{array}{lll}
\hbox{type A}: & I (p,0;r;y_A) \, ,
       & y_A = -\frac{4}{3}p+r \, ;\\[1ex]
\hbox{type B}: & I (p,0;r;y_B) \, ,
       & y_B = -\frac{4}{3}p-r-2 \, ;\\[1ex]
\hbox{type C}: & I (0,q;r;y_C) \, ,
       & y_C = \frac{4}{3}q+r+4 \, ;\\[1ex]
\hbox{type D}: & I (0,q;r;y_D) \, ,
       & y_D = \frac{4}{3}q-r+2 \, , \,
       \hbox{ and } (q,r) \neq (0,0), \, .
  \end{array}
\end{eqnarray*}

\end{Theorem}

We shall construct below certain $E(3,8)$-morphisms between the modules
$M(p,q;r,y_X)$.  This will imply that all
modules $I(p,q;r,y)$ on the list are degenerate.  The proof of
the fact that
the list is complete will be published elsewhere.\\

The theorem means that all the degenerate generalized Verma modules over
$E(3,8)$ are in fact the direct summands of induced modules
$M(V_X)$, $X=A,B,C,D$:
\begin{eqnarray*}
  M(V_X) &=& \bigoplus_{m,n \in \ZZ} M(V^{m,n}_X),
     \hbox{ where}\\
V^{m,n}_X &=& \left\{ f\,\1_X \,|\,\,\, (\sum x_i\partial_i)
  f=mf, \,\,(\sum z_a\partial_a) f=n  f \right\} \, ,
\end{eqnarray*}
(we normalize degree of $\1_X$ as $(0,0)$).

We construct morphisms between these modules with
the help of Proposition~\ref{prop:1.1}.  As in the $E^{\flat}$
case \cite{KR2}, introduce the following operators on $M (\Hom
(V_X,V_X))$:
\begin{eqnarray*}
  \triangledown &=& \Delta^+ \delta_+ + \Delta^- \delta_-
     = \delta_1 \partial_1 + \delta_2\partial_2 + \delta_3\partial_3 \,
     , \hbox{ where}\\
\Delta^{\pm} &=& \sum^3_{i=1} d^{\pm}_i \otimes \partial_i \, ,
\, \delta_i =\sum_{\alpha = \pm} d^a_i \otimes \partial_a \, .
\end{eqnarray*}

\begin{Proposition}
  \label{prop:3.2} (a)~~The element $\triangledown$ gives a
  well-defined morphism $M(V_X) \to M(V_X)$, \break$X=A,B,C,D$, by
  formula~(\ref{eq:1.2}).

(b)~~$\triangledown^2 =0 $.

\end{Proposition}

\begin{proof}
  The proof of (b) is the same as in \cite{KR2}.  In order to
  prove~(a), we have to check conditions (\ref{eq:1.4a}) and
  (\ref{eq:1.4b}).  It is obvious that $\Delta^{\pm}$
  (resp. $\delta_i$) are $s\ell_3(\CC)$-(resp. $s\ell_2
  (\CC)$-) invariant.  Using both formulas for $\triangledown$, we conclude
  that it is $\fg_0$-invariant, proving (\ref{eq:1.4a}).  In order
    to check (\ref{eq:1.4b}), first note that
    \begin{displaymath}
      e'_0\triangledown =
                 (f_2\partial_1\partial_+ - f_{12}\partial_2\partial_+)
        = (x_3\partial_2\partial_1-x_{3}\partial_1\partial_2)\,\partial_+
        =0 \, .
    \end{displaymath}
Now
\begin{eqnarray*}
  e^{\sharp}_0\triangledown &=& h^{\sharp}_0 \partial_1\partial_+
    + f_1\partial_2\partial_+ + f_{12}\partial_3\partial_1
    - f_3\partial_1\partial_- \\
    &=& (x_1 \partial_1 - z_+ \partial_+ +T)\, \partial_1 \partial_+
        + x_2\partial_1\partial_2\partial_+
        +x_3\partial_1\partial_3\partial_+
        -z_- \partial_+\partial_1\partial_-\\
    &=& (x_1 \partial_1 - z_+ \partial_+ +T +x_2\partial_2
       + x_3\partial_3 - z_- \partial_-)\, \partial_1 \partial_+ =0 \, ,
\end{eqnarray*}
where we use (\ref{eq:3.6}) to check that
\begin{equation}
  \label{eq:3.7}
  h^{\sharp}_0 = \tfrac{2}{3}h_1 + \tfrac{1}{3} h_2
     -\tfrac{1}{2} h_3 + \tfrac{1}{2} Y^{\sharp}
                   = x_1\partial_1-z_+\partial_+  +T \, .
\end{equation}

\end{proof}

Let $M_{X'} =M(V_{X'})$, where:
\begin{eqnarray}
  \label{eq:3.8}
  V_{A'} &=& \CC [x_1,x_2,x_3] \1_A =\oplus_{\,p \geq 0}
     V^{p, 0}_A  \, , \,\,
     V_{B'} = \CC [x_{{1}},x_2,x_3]\1_B
     =\oplus_{p\geq 0} V^{\,p, 0}_B \, \\
\nonumber
     V_{C'} &=& \CC [\partial_1,\partial_2,\partial_3]  \1_C
     = \oplus_{q \geq 0} V^{-q,0}_C \, , \,
     V_{D'} = \CC [\partial_1,\partial_2,\partial_3] \1_D
     =\oplus_{q \geq 0} V^{-q,0}_D
\end{eqnarray}
and let
\begin{equation}
  \label{eq:3.9}
  \triangledown_2 = \Delta^- \Delta^+ = d^-_1 \Delta^+ \partial_1
   + d^-_2\Delta^+\partial_2 +d^-_3 \Delta^+\partial_3 \, .
\end{equation}
We define the morphism $\triangledown_2 : M_{A'} \to M_{B'}$ by extending
the map defined by $\triangledown_2$ as follows
\begin{eqnarray}
\nonumber
 && V_{A'} \overset{\sim}{\rightarrow} \CC [x_1,x_2,x_3]
     \longrightarrow U(L_-) \otimes \CC [x_1,x_2,x_3]
     \1_B \simeq U(L_-) \otimes V_{B'} \, , \\
  \label{eq:3.10}
&&\,\,  f \1_A \,\,\mapsto\quad f
 \quad  \qquad\mapsto\qquad\quad \triangledown_2  f  \1_B \, .
\end{eqnarray}
In order to apply Proposition~\ref{prop:1.1}, we have to check
conditions (\ref{eq:1.4a}) and (\ref{eq:1.4b}) for
$\triangledown_2$.  As before, condition~(\ref{eq:1.4a})
obviously holds.  Now
\begin{displaymath}
  e'_0 \triangledown_2  \1_B =-\sum_i d^-_i
  (f_2\partial_1\partial_- - f_{12}\partial_2) \partial_i f  \1_B =0
\end{displaymath}
because $f_2 \partial_1 - f_{12} \partial_2 = x_3 \partial_2 \partial_1 -
x_3\partial_1\partial_2=0$.  Furthermore:
\begin{eqnarray*}
  e^{\sharp}_0 \triangledown_2 f  \1_B
  &=&   \left(-\sum_i d^-_i (h^{\sharp}_0 \partial_1
      + f_1\partial_2  +f_{12}\partial_3) \partial_i -f_3 \Delta^+
        \partial_1\right) f \1_B \\
  &=& -\left(\sum_i d^-_i (x_1\partial_1-z_+\partial_+ +T
  +x_2\partial_2+x_3\partial_3)
  \partial_1 \partial_2-\Delta^-\partial_1
    - \Delta^+\partial_1 f_3 \right) f  \1_B \, .
\end{eqnarray*}
As $z_{\pm} \partial_{\pm}  f \1_B = f
z_{\pm}\partial_{\pm}  \1_B =f
(-1+\partial_{\pm}z_{\pm})\1_B=-f\1_B$, and $f_3 f \1_B =0$, we conclude that
\begin{eqnarray*}
  -e^{\sharp}_0\triangledown_2 f\1_B  &=& \sum_i d^-_i
     (x_1\partial_1 +x_2\partial_2 +x_3\partial_3 -z_+\partial_+ +T +1)
     f\1_B  \qquad\qquad\qquad\qquad\qquad\quad\\
          &=&  \sum_i d^-_i (z_- \partial_- +1) f \1_B =0 \, .
\end{eqnarray*}
Thus Proposition~\ref{prop:1.1} applies and we get the morphism
$\triangledown_2 : M_{A'} \to M_{B'}$.

In exactly the same fashion we construct the morphism
$\triangledown_2 : M_{C'} \to M_{D'}$ by taking $f \in \CC
[\partial_1,\partial_2, \partial_3]$.

Thus we have proved part (a) of the following proposition;
part~(b) was checked in \cite{KR2}.

\begin{Proposition}
  \label{prop:3.3} (a)~~Formulae (\ref{eq:3.9}) and
  (\ref{eq:3.10}) define the morphisms $\triangledown_2 :M_{A'}
  \to M_{B'}$, $\triangledown_2 : M_{C'} \to M_{D'}$ of
  $E(3,8)$-modules.

(b)~~$\triangledown\triangledown_2 =0$, $\triangledown_2
\triangledown =0$.
\end{Proposition}

Similarly we can construct morphisms with the help of the element
\begin{equation}
  \label{eq:3.11}
  \triangledown_3 = \delta_1\delta_2\delta_3 =\sum_{a,b,c}
  d^a_1 d^b_2d^c_3 \otimes \partial_a\partial_b\partial_c \,,\, \,\,
  \hbox{\footnotesize{$(a,b,c=+,-)$}} \, .
\end{equation}
Let us consider modules $M_{X''} = M(V_{X''})$, where
\begin{eqnarray*}
  V_{A''} &=& \CC [z_+,z_-]  \1_A = \oplus_{r \geq 0}
        V^{0,r}_A \, , \,\,\,
        V_{B''} \,=\, \CC [\partial_+,\partial_-] \1_B
        = \oplus_{r\geq 0} V^{0,-r}_B \, ,\\[1ex]
  V_{C''} &=& \CC [z_+,z_-]  \1_C = \oplus_{r \geq 0}
        V^{0,r}_B \, ,\,\, \,
        V_{D''}\, = \,\CC [\partial_+,\partial_-]  \1_D
        = \oplus_{r \geq 0} V^{0,-r}_D \, .
\end{eqnarray*}
We construct morphisms $\triangledown_3 : M_{A''} \to M_{C''}$
and $\triangledown_3 : M_{B''} \to M_{D''}$ by extending the maps
$\triangledown'_3 : V_{A''} \to U(L_-) \otimes V_{C''}$,
$\triangledown''_3 : V_{B''} \to U(L_-) \otimes V_{D''}$ given by the left and
right diagrams below:
\begin{eqnarray}      %
\label{eq:3.12}\hspace*{.3in}
\raisebox{4.5ex}{$\CC [z_a] \cdot \1_A \overset{\sim}{\longrightarrow}$}
\begin{CD}
   \CC [z_+,z_-]\\
   @VV{\triangledown_3}V\\
    U(L_-) \otimes (\CC [z_+,z_-] \1_C)
\end{CD}
\hspace{.3in}
\raisebox{4.5ex}{$\CC [\partial_a] \cdot \1_B
     \overset{\sim}{\longrightarrow}$}
\begin{CD}
   \CC [\partial_+,\partial_-]\\
   @VV{\triangledown_3}V\\
    U(L_-) \otimes (\CC [\partial_+,\partial_-] \1_D) \, .
\end{CD}
\end{eqnarray}
Here the horizontal maps are naturally $\,s\ell_3 (\CC) \oplus s\ell_2
(\CC)$-isomorphisms, but we have to define the action of $Y$ on
the target demanding the map to be a $\fg_0$-isomorphism.  With
this in mind we have to check that $Y$ commutes with
$\triangledown_3$.

We shall use the
Einstein summation convention argument for the vertical maps
$\triangledown_3$ given by (\ref{eq:3.12}).  Then for
$f \in \CC[z_+,z_-]$ we have $Y (f \1_A) = (\deg f) f \1_A$ and
\begin{eqnarray*}
  Y (d^a_1d^b_2d^c_3 \otimes \partial_a\partial_b\partial_c f \1_C)
  &=& d^a_1d^b_2d^c_3 \otimes (-1+Y^{\sharp})
     \partial_1\partial_b\partial_c f \1_C\\
  &=& d^a_1d^b_2d^c_3 \otimes \partial_a\partial_b\partial_c f
     (-4+\deg f +Y^{\sharp}) \1_C \\
 &=& (\deg f) \cdot d^a_1 d^b_2d^c_3 \otimes
       \partial_a\partial_b\partial_c f \1_C \, .
\end{eqnarray*}
Similarly for $f \in \CC[\partial_+,\partial_-]$, we have
$Y (f \1_B) = (-\deg f-2)f\1_B$ and
\begin{eqnarray*}
  Y(d^a_1 d^b_2d^c_3 \otimes \partial_a\partial_b\partial_c f\1_D)
    &=& d^a_1 d^b_2d^c_3 \otimes \partial_a\partial_b\partial_c f
    (-4-\deg f +Y^{\sharp}) \1_D \\
    &=& (-4-\deg f+2) d^a_1d^b_2d^c_3 \otimes
        \partial_a\partial_b\partial_c f \1_D \, ,
\end{eqnarray*}
where $Y^{\sharp}$ defined by (\ref{eq:3.6}). Thus we get the commutativity.

One meets no problem checking $e'_0 \triangledown_3=0$, but we consider calculations for
$e^{\sharp}_0 \triangledown_3$ in more detail.  If $f \in
\CC [z_+,z_-]$, then
\begin{displaymath}
  e_0(\triangledown_3 f\1_C) = h^{\sharp}_0
  d^b_2d^c_3 \otimes \partial_+\partial_b\partial_c f\1_C -
  f_3 d^b_2d^c_3 \otimes \partial_-\partial_b\partial_c f\1_C \, ,
\end{displaymath}
because $d^a_1 (x_2\partial_1) d^c_3 \otimes
\partial_a\partial_+\partial_c f\1_C = d^a_1d^b_2 (x_3\partial_1)
\otimes \partial_a\partial_b\partial_+ f\1_C=0$.
Now
$  f_3 (d^b_2d^c_3 \otimes \partial_b\partial_c)
  = (d^b_2d^c_3 \otimes \partial_b\partial_c)f_3$
and
$h^{\sharp}_0(d^b_2d^c_3 \otimes \partial_b\partial_c) = (d^b_2d^c_3 \otimes
\partial_b\partial_c) (h^{\sharp}_0-2)$,
where $h^{\sharp}_0 = \tfrac{2}{3}h_1 + \tfrac{1}{3}h_2 -
\tfrac{1}{2}h_3 + \tfrac{1}{2}Y^{\sharp}$,
and again $Y^{\sharp}$ is defined by (\ref{eq:3.6}).  Therefore
\begin{eqnarray*}
  e^{\sharp}_{{0}} (\triangledown_3 f\1_C)
  &=& (d^b_2d^c_3 \otimes \partial_b\partial_c)
    (h^{\sharp}_{{0}} \partial_+-2\partial_+-z_+\partial_+\partial_-)
       f\1_C \\
  &=& (d^b_2d^c_3 \otimes \partial_b\partial_c)
     (h^{\sharp}_{{0}} -z_-\partial_- -2) \partial_+ f\1_C\\
  &=& (d^b_2d^c_3 \otimes \partial_b\partial_c)
     (-x_2\partial_2-x_3\partial_3-2) \partial_+ f \1_C \\
  &=& (d^b_2d^c_3 \otimes \partial_b\partial_c)
     (-\partial_2x_2 - \partial_3x_3) \partial_+ f\1_C =0 \, .
\end{eqnarray*}

The calculations in the case $f \in \CC [\partial_+,\partial_-]$ are very much the
same.  So we have proved part~(a) of the following proposition;
part~(b) was checked in \cite{KR2}

\begin{Proposition} (a)~~Formulae (\ref{eq:3.11}) and
  (\ref{eq:3.12}) define the morphisms $\triangledown_3 :M_{A''}
  \to M_{D''}$ and $M_{B''} \to M_{D''}$ of $E(3,8)$-modules.

(b)~~$\triangledown \cdot \triangledown_3 =0 \, , \quad
  \triangledown_3 \cdot \triangledown =0 $.
\end{Proposition}

Furthermore, there are $E(3,8)$-module morphisms
\begin{eqnarray*}
  \triangledown'_4 &:& M (00;2;y_A) \to
  M (01;1;y_D) \hbox{ and }\\
  \triangledown''_4 &:& M (10;0;y_A) \to
  M(00;2;y_D)
\end{eqnarray*}
defined by formulae (2.14) and (2.17) from \cite{KR2}, applied to
$E(3,8)$.  Arguments similar to those in \cite{KR2} show that
these are indeed well-defined morphisms.

Thus far we have constructed $E(3,8)$-homomorphisms
$\triangledown$, $\triangledown_2$, $\triangledown_3$,
$\triangledown'_4$, and $\triangledown''_4$ between generalized
Verma modules.  Note that these maps have degree $1$, $2$, $3$
and $4$, respectively with respect to the $\ZZ$-gradation of
$U(L_-)$ induced by that of $E(3,8)$.

As in the case of $E(3,6)$
\cite{KR2}, all these maps are illustrated in Figure~1.  The
nodes in the quadrants A,B,C,D represent generalized Verma
modules $M(p,0;r;y_X)$ if $X=A$ or $B$,
and $M(0,q;r;y_X) $, if $X=C$ or $D$.  The plain arrows
represent $\triangledown$, the dotted arrows represent
$\triangledown_2$, the interrupted arrows represent
$\triangledown_3$ and the bold arrows represent
$\triangledown'_4$ and $\triangledown''_4$.


\begin{figure}[htbp]
  \begin{center}
    \leavevmode
  \setlength{\unitlength}{0.25in}
\begin{picture}(21,19)

\put(18.7,17.5){A}
\put(18.7,-1){B}

\put(1,17.5){C}
\put(1,-1){D}

\put(8.5,17.5){\line(0,-1){7} }
\put(8.4,18){r}

\put(11.5,17.5){\line(0,-1){8.5} }
\put(11.4,18){r}

\put(10,-.5){\line(0,1){9.5} }
\put(9.9,-1){r}

\put(13,-.5){\line(0,1){8} }
\put(12.9,-1){r}

\put(13,7.5){\line(1,0){5} }
\put(18.5,7.4){p}

\put(1.8,9){\line(1,0){8.2} }
\put(18.5,8.9){p}

\put(11.5,9){\line(1,0){6.5} }
\put(1,8.9){q}

\put(1.8,10.5){\line(1,0){6.8} }
\put(1,10.4){q}

\thicklines
\multiput(2.5,0)(0,1.5){12}{\circle{.25} }
\multiput(4,0)(0,1.5){12}{\circle{.25} }
\multiput(5.5,0)(0,1.5){12}{\circle{.25} }
\multiput(7,0)(0,1.5){12}{\circle{.25} }

\multiput(8.5,0)(0,1.5){12}{\circle{.25} }

\multiput(10,0)(0,1.5){7}{\circle{.25} }

\multiput(11.5,9)(0,1.5){6}{\circle{.25} }

\multiput(13,0)(0,1.5){12}{\circle{.25} }

\multiput(14.5,0)(0,1.5){12}{\circle{.25} }
\multiput(16,0)(0,1.5){12}{\circle{.25} }
\multiput(17.5,0)(0,1.5){12}{\circle{.25} }

\thinlines

\thinlines
\drawline(1.95,-.55)(2.4,-.1)
\drawline(1.95, .95)(2.4,1.4)
\drawline(1.95,2.45)(2.4,2.9)
\drawline(1.95,3.95)(2.4,4.4)
\drawline(1.95,5.45)(2.4,5.9)
\drawline(1.95,6.95)(2.4,7.4)
\drawline(1.95,8.45)(2.4,8.9)

\drawline(1.95,11.45)(2.4,11.9)
\drawline(1.95,12.95)(2.4,13.4)
\drawline(1.95,14.45)(2.4,14.9)
\drawline(1.95,15.95)(2.4,16.4)

\drawline(3.45,-.55)(3.9,-.1)
\drawline(4.95,-.55)(5.4,-.1)
\drawline(6.45,-.55)(6.9,-.1)
\drawline(7.95,-.55)(8.4,-.1)
\drawline(9.45,-.55)(9.9,-.1)

\drawline(13.95,-.55)(14.4,-.1)
\drawline(15.45,-.55)(15.9,-.1)
\drawline(16.95,-.55)(17.4,-.1)

\multiput(18.2,.7)(0,1.5){12}{\vector(-1,-1){.55}}
\multiput(3.2,17.2)(1.5,0){5}{\vector(-1,-1){.55}}
\multiput(12.2,17.2)(1.5,0){5}{\vector(-1,-1){.55}}


\multiput(3.85,16.35)(1.5,0){4}{\vector(-1,-1){1.2}}
\multiput(12.85,16.35)(1.5,0){4}{\vector(-1,-1){1.2}}

\multiput(3.85,14.85)(1.5,0){4}{\vector(-1,-1){1.2}}
\multiput(12.85,14.85)(1.5,0){4}{\vector(-1,-1){1.2}}

\multiput(3.85,13.35)(1.5,0){4}{\vector(-1,-1){1.2}}
\multiput(12.85,13.35)(1.5,0){4}{\vector(-1,-1){1.2}}

\multiput(3.85,11.85)(1.5,0){4}{\vector(-1,-1){1.2}}
\multiput(12.85,11.85)(1.5,0){4}{\vector(-1,-1){1.2}}
\dashline[-10]{.1}(3.85,10.35)(2.35,8.85)
       \put(7.33,9.37){\vector(-1,-1){.25}}
\dashline[-10]{.1}(5.35,10.35)(3.85,8.85)
       \put(5.8,9.37){\vector(-1,-1){.25}}
\dashline[-10]{.1}(6.85,10.35)(5.35,8.85)
       \put(4.3,9.37){\vector(-1,-1){.25}}
\dashline[-10]{.1}(8.35,10.35)(6.85,8.85)
       \put(2.9,9.35){\vector(-1,-1){.25}}
\multiput(12.85,10.35)(1.5,0){4}{\vector(-1,-1){1.2}}
\multiput(3.85,8.85)(1.5,0){5}{\vector(-1,-1){1.2}}
\dashline[-10]{.1}(14.35,8.85)(12.85,7.35)
\dashline[-10]{.1}(15.85,8.85)(14.35,7.35)
\dashline[-10]{.1}(17.35,8.85)(15.85,7.35)
\dashline[-10]{.1}(2.25,10.25)(2,10)
      \put(13.4,7.9){\vector(-1,-1){.25}}
      \put(14.9,7.9){\vector(-1,-1){.25}}
      \put(16.4,7.9){\vector(-1,-1){.25}}

\dashline[-10]{.2}(10.3,16.8)(8.5,15) 
       \put(8.8,15.35){\vector(-1,-1){.25}}

\dashline[-10]{.2}(11.5,16.5)(8.5,13.5)    
       \put(8.8,13.85){\vector(-1,-1){.25}}

\dashline[-10]{.2}(11.5,15)(8.5,12)
       \put(8.8,12.35){\vector(-1,-1){.25}}

\dashline[-10]{.2}(11.5,13.5)(8.5,10.5)
       \put(8.8,10.85){\vector(-1,-1){.25}}

\thicklines
\put(11.5,12){\vector(-1,-1){2.81}}
\put(13,9){\vector(-1,-1){2.81}}


\thinlines


\dashline[-10]{.2}(13,7.5)(10,4.5)
      \put(10.4,4.9){\vector(-1,-1){.25}}

\dashline[-10]{.2}(13,6)(10,3)
      \put(10.4,3.4){\vector(-1,-1){.25}}

\dashline[-10]{.2}(13,4.5)(10,1.5)
      \put(10.4,1.9){\vector(-1,-1){.25}}

\dashline[-10]{.2}(13,3)(10,0)
     \put(10.4,.4){\vector(-1,-1){.25}}

\dashline[-10]{.2}(13,1.5)(11,-.5)
\dashline[-10]{.2}(13,0)(12.5,-.5)

\multiput(3.85,7.35)(1.5,0){5}{\vector(-1,-1){1.2}}
\multiput(14.35,7.35)(1.5,0){3}{\vector(-1,-1){1.2}}
\multiput(3.85,5.85)(1.5,0){5}{\vector(-1,-1){1.2}}
\multiput(14.35,5.85)(1.5,0){3}{\vector(-1,-1){1.2}}
\multiput(3.85,4.35)(1.5,0){5}{\vector(-1,-1){1.2}}
\multiput(14.35,4.35)(1.5,0){3}{\vector(-1,-1){1.2}}
\multiput(3.85,2.85)(1.5,0){5}{\vector(-1,-1){1.2}}
\multiput(14.35,2.85)(1.5,0){3}{\vector(-1,-1){1.2}}
\multiput(3.85,1.35)(1.5,0){5}{\vector(-1,-1){1.2}}
\multiput(14.35,1.35)(1.5,0){3}{\vector(-1,-1){1.2}}

\end{picture}
\vspace{3ex}
    \caption{}
    \label{fig:1}
  \end{center}
\end{figure}
Note that the generalized Verma modules $M(00;1;y_A)$ and
$M(00;1;y_D)$ are isomorphic since $y_A = y_D =1$.  We shall
identify them.  This allows us to construct the $E(3,8)$-module
homomorphism
\begin{displaymath}
  \tilde{\triangledown} :
     M(00;1;y_A) \to M(01;2;y_D)\, ,
\end{displaymath}
    which is NOT represented in Figure~1.

Note that $I (00;1;y_A) = I(00;1;y_D)$ is the coadjoint
$E(3,8)$-module.  It follows from the above propositions that
if we remove the module $M(00;1;y_D)$ from
Figure~1, and draw $\tilde{\triangledown}$
then all sequences in the modified Figure~1 become complexes.
We denote by $H^{p,-r}_A$, $H^{p,r}_B$, $H^{-q,r}_C$
and $H^{-q,-r}_D$ the homology of these complexes
at the position of $M(pq;r;y_X)$, $X=A,B,C,D$.

\begin{Theorem}
  \label{th:3.5}
  (a)~~The kernels of all maps $\triangledown$,  $\triangledown_2$,
  $\triangledown_3$,  $\triangledown'_4$,  $\triangledown''_4$,
  $\tilde{\triangledown}$ are maximal submodules.

  (b)~~The homology $H^{m,n}_X$ is zero except for six cases
  listed (as $E(3,8)$-modules) below:
  \begin{eqnarray*}
 \hspace*{-5ex}H^{0,0}_A &=& \CC \, ,\,\,\,
    \quad H^{1,1}_A \,\,=\,\, I(10;0;-\tfrac{4}{3}) \, ,\\[1ex]
    H^{1,0}_A &=& H^{0,-2}_D\,\, =\,\, I(00;0;-2) \, , \\[1ex]
    H^{-1,-1}_D &=& H^{-1,-2}_D \,\,=\,\,  I(00;1;1) \oplus \CC \, .
\end{eqnarray*}

\end{Theorem}

The proof is similar to that of the analogous $E(3,6)$-result in
\cite{KR2}.  Note that this theorem gives the following explicit
construction of all degenerate irreducible $E(3,8)$-modules:
\begin{displaymath}
  I(pq;r;y_X) = M(pq;r;y_X)/\Ker \triangledown \, ,
\end{displaymath}
where $\triangledown$ is the corresponding map in the modified
Figure~1.


\section{Three series of degenerate Verma modules over $E(5,10)$}

As in \cite{KR2} and in \S~\ref{sec:2}, we use for the odd
elements of $E(5,10)$ the notation $d_{ij} = dx_i \wedge dx_j$
$(i,j=1,2,\ldots ,5)$; recall that we have the following
commutation relation $(f,g \in \CC [[ x_1 , \ldots , x_5 ]])$:
\begin{displaymath}
  [f d_{jk}, gd_{\ell m}] = \epsilon_{ij k \ell m} \partial_i
  \, ,
\end{displaymath}
where $\epsilon_{ijk\ell m}$ is the sign of the permutation
$ijk\ell m$ if all indices are distinct and $0$ otherwise.

Recall that the Lie superalgebra $E(5,10)$ carries a unique
consistent irreducible $\ZZ$-gradation $E(5,10) = \oplus_{j \geq
  -2}\, p_j$.  It is defined by:
\begin{displaymath}
  \deg x_i =2 =-\deg \partial_i \, , \,\, \deg d_{ij} =-1 \, .
\end{displaymath}
One has:  $p_0 \simeq s\ell_5 (\CC)$ and the $p_0$-modules
occurring in the $L_-$ part are:
\begin{eqnarray*}
  p_{-1} &=& \langle d_{ij} \,\, | i,j=1,\ldots ,5 \rangle \simeq
     \Lambda^2 \CC^5 \, ,\\
     p_{-2} &=& \langle \partial_i \,\, | i=1,\ldots ,5 \rangle
        \simeq \CC^{5*} \, .
\end{eqnarray*}
Recall also that $p_1$ consist of closed $2$-forms with linear
coefficients, that $p_1$ is an irreducible $p_0$-module and $p_j
= p^j_1$ for $j \geq 1$.

We take for the Borel subalgebra of $p_0$ the subalgebra of the
vector fields $\langle \sum_{i \leq j} a_{ij} x_i \partial_j |
a_{ij} \in \CC \, , \, \tr (a_{ij})=0 \rangle$, and denote by
$F(m_1,m_2,m_3,m_4)$ the finite-dimensional irreducible
$p_0$-module with highest weight $(m_1,m_2,m_3,m_4)$.  We let
\begin{displaymath}
  M(m_1,m_2,m_3,m_4) = M(F(m_1,m_2,m_3,m_4))
\end{displaymath}
denote the corresponding generalized Verma module over $E(5,10)$.

\begin{Conjecture}
  \label{conj:4.1}
The following is a complete list of generalized Verma modules
over $E(5,10)$  $(m,n \in \ZZ_+)$:
\begin{displaymath}
  M(m,n,0,0)\, , \, M(0,0,m,n) \hbox{ and } M(m,0,0,n) \, .
\end{displaymath}

\end{Conjecture}

In this section we construct three complexes of generalized
$E(5,10)$ Verma modules which shows, in particular, that all
modules from the list given by Conjecture~\ref{conj:4.1} are
degenerate.  We let:
\begin{displaymath}
  S_A = S (\CC^5 + \Lambda^2 \CC^5) \, , \,
  S_B = S(\CC^{5*} + \Lambda^2 \CC^{5*}) \, , \,
  S_C = S (\CC^5 + \CC^{5*}) \, .
\end{displaymath}
Denote by $x_i$ $(i=1,\ldots ,5)$ the standard basis of $\CC^5$
and by $x_{ij}=-x_{ji}$ $(i,j=1, \ldots ,5)$ the standard
basis of $\Lambda^2 \CC^5$.  Let $x^*_i$ and $x^*_{ij} =-x^*_{ji}$
be the dual bases of $\CC^{5*}$ and $\Lambda^2\CC^{5*}$,
respectively.  Then $S_A$ is the polynomial algebra in
$15$~indeterminates $x_i$ and $x_{ij}$, $S_B$ is the polynomial
algebra in $15$~indeterminates $x^*_i$ and $x^*_{ij}$ and $S_C$
is the polynomial algebra in $10$~indeterminates $x_i$ and
$x^*_i$.

Given two irreducible $p_0$-modules $E$ and $F$, we denote by $(E
\otimes F)_{\high}$ the highest irreducible component of the
$p_0$-module $E \otimes F$.  If $E = \oplus_i E_i$ and $F=
\oplus_jF_j$ are direct sums of irreducible $p_0$-modules, we let
$(E \otimes F)_{\high} = \oplus_{i,j} (E_i \otimes F_j)_{\high}$.
If $E$ and $F$ are again irreducible $p_0$-modules, then $S (E
\oplus F)= \oplus_{m,n \in \ZZ_+} S^m E \otimes S^nF$, and we let
$S_{\high} (E \oplus F) = \oplus_{m,n \in \ZZ_+} (S^m E \otimes
S^nF)_{\high}$.  We also denote by $S_{\low} (E \oplus F)$ the
complement to $S_{\high}(E \oplus F)$.

It is easy to see that we have as $p_0$-modules:
\begin{eqnarray*}
  S_{A,\high} & \simeq & \oplus_{m,n \in \ZZ_+}
      F (m,n,0,0) \, , \\
  S_{B,\high}  & \simeq & \oplus_{m,n \in \ZZ_+}
      F(0,0,m,n) \, , \\
  S_{C,\high}  & \simeq & \oplus_{m,n \in \ZZ_+}
      F (m,0,0,n) \, .
\end{eqnarray*}

Introduce the following operators on the spaces $M(\Hom
(S_X,S_X))$, $X=A$, $B$ or $C$:
\begin{displaymath}
  \triangledown_X = \sum^5_{i,j=1} d_{ij} \otimes \theta^X_{ij}
  \, ,
\end{displaymath}
where
\begin{displaymath}
  \theta^A_{ij} = \frac{d}{dx_{ij}} \, , \,
  \theta^B_{ij} = x^*_{ij} \, , \,
  \theta^C_{ij} =x^*_i \frac{d}{dx_j}-x^*_j\frac{d}{dx_i} \, .
\end{displaymath}

It is immediate to see that $p_0 \cdot \triangledown_X =0$.  In
order to apply Proposition~\ref{prop:1.1}, we need to check that
\begin{equation}
  \label{eq:4.1}
  p_1 \cdot \triangledown_X =0 \, .
\end{equation}
This is indeed true in the case $X=C$, but it is not true in the
cases $X=A$ and $B$.  In fact (\ref{eq:4.1}) applied to $f \in
S_X$, $X=A$ or $B$, is equivalent to the following equations,
respectively $(a,b,c,d=1,\ldots,5)$:
\begin{eqnarray}
  \label{eq:4.2}
\left( \frac{d}{dx_{ab}}\frac{d}{dx_{cd}}
     - \frac{d}{dx_{ac}}\frac{d}{dx_{bd}}
     + \frac{d}{dx_{ad}}\frac{d}{dx_{bc}}
\right) f & =& 0\, , \\ [2ex]
  \label{eq:4.3}
  \left(
    x^*_{ab} x^*_{cd}
      -   x^*_{ac} x^*_{bd}
      +   x^*_{ad} x^*_{bc}
\right) f & = & 0 \, .
\end{eqnarray}
It is not difficult to check the following lemma.

\begin{Lemma}
  \label{lem:4.2}
(a)~~The subspace of $S_A$ defined by equations (\ref{eq:4.2}) is
$S_{A,\high}$.

(b)~~Equations (\ref{eq:4.3}) hold in $S_B / S_{B,\low}$.

(c)~~Equation $\triangledown^2_X =0$ is equivalent to the system
of equations $(a,b,c,d=1,\ldots ,5)$:
\begin{displaymath}
  \theta_{ab} \theta_{cd}-\theta_{ac}\theta_{bd}
  + \theta_{ad}\theta_{bc} =0 \, .
\end{displaymath}

\end{Lemma}

We let:
\begin{displaymath}
  V_A = S_{A,\high} \, , \, V_{B (\hbox{resp. }C)} =
  S_{B  (\hbox{resp. }C)}/ S_{B  (\hbox{resp. }C)_{,\low}}
\end{displaymath}
The above discussion implies

\begin{Proposition}
  \label{prop:4.3}
(a)~~The operators $\triangledown_X$ define $E(5,10)$-morphisms
$M(V_X) \to M(V_X) \,\, (X=A,B \hbox{ or } C)$.

(b)~~$\triangledown^2_X=0 \,\, (X=A,B \hbox{ or } C)$.

(c)~~$\triangledown_X = 0 $ iff $X=A$ and $n=0$,  or
$X=C$ and $m=0$.
\end{Proposition}

The non-zero maps $\triangledown_X$ are illustrated in
Figure~\ref{fig:2}.  The nodes in the quadrants $A$, $B$ and $C$
represent generalized Verma modules $M(m,n,0,0)$, $M(0,0,m,n)$
and $M(m,0,0,n)$, respectively.  The arrows represent the
$E(5,10)$-morphisms $\triangledown_X$, $X = A$, $B$ or $C$ in the
respective quadrants.

\begin{figure}[htbp]
  \begin{center}
    \leavevmode
  \setlength{\unitlength}{0.25in}
\begin{picture}(21,19)


\put(18.7,17.5){A (m,n,0,0)}
\put(18.7,-1){C (m,0,0,n)}

\put(1,17.5){}
\put(1,-1){B (0,0,m,n)}


\put(10,17.5){\line(0,-1){8.5} }
\put(9.9,18){n}

\put(10,-.5){\line(0,1){9.5} }
\put(9.9,-1){n}



\put(1.8,9){\line(1,0){8.2} }
\put(1,8.9){m}

\put(10,9){\line(1,0){8} }
\put(18.5,8.9){m}


\thicklines
\multiput(2.5,0)(0,1.5){7}{\circle{.25} }
\multiput(4,0)(0,1.5){7}{\circle{.25} }  
\multiput(5.5,0)(0,1.5){7}{\circle{.25} }
\multiput(7,0)(0,1.5){7}{\circle{.25} }  
\multiput(8.5,0)(0,1.5){7}{\circle{.25} }

\multiput(10,0)(0,1.5){12}{\circle{.25} }  
\multiput(11.5,0)(0,1.5){12}{\circle{.25} }
\multiput(13,0)(0,1.5){12}{\circle{.25} }  
\multiput(14.5,0)(0,1.5){12}{\circle{.25} }
\multiput(16,0)(0,1.5){12}{\circle{.25} }  
\multiput(17.5,0)(0,1.5){12}{\circle{.25} }

\thinlines

\drawline(1.95,0  )(2.35,0  )
\drawline(1.95,1.5)(2.35,1.5)
\drawline(1.95,3  )(2.35,3  )
\drawline(1.95,4.5)(2.35,4.5)
\drawline(1.95,6  )(2.35,6  )
\drawline(1.95,7.5)(2.35,7.5)
\drawline(1.95,9  )(2.35,9  )


\drawline(10.95,-.55)(11.4,-.1)
\drawline(12.45,-.55)(12.9,-.1)
\drawline(13.95,-.55)(14.4,-.1)
\drawline(15.45,-.55)(15.9,-.1)
\drawline(16.95,-.55)(17.4,-.1)

\multiput(18.2,.7)(0,1.5){5}{\vector(-1,-1){.55}}
\multiput(10,17.25)(1.5,0){5}{\vector(0,-1){.55}}


\multiput(10,16.3)(1.5,0){6}{\vector(0,-1){0.9}}
\multiput(10,14.8)(1.5,0){6}{\vector(0,-1){0.9}}
\multiput(10,13.3)(1.5,0){6}{\vector(0,-1){0.9}}
\multiput(10,11.8)(1.5,0){6}{\vector(0,-1){0.9}}
\multiput(10,10.3)(1.5,0){6}{\vector(0,-1){0.9}}

\multiput(3.7,9  )(1.5,0){5}{\vector(-1,0){0.9}}
\multiput(3.7,7.5)(1.5,0){5}{\vector(-1,0){0.9}}
\multiput(3.7,6  )(1.5,0){5}{\vector(-1,0){0.9}}
\multiput(3.7,4.5)(1.5,0){5}{\vector(-1,0){0.9}}
\multiput(3.7,3  )(1.5,0){5}{\vector(-1,0){0.9}}
\multiput(3.7,1.5)(1.5,0){5}{\vector(-1,0){0.9}}
\multiput(3.7,0  )(1.5,0){5}{\vector(-1,0){0.9}}

\multiput(11.35,8.85)(1.5,0){5}{\vector(-1,-1){1.2}}
\multiput(11.35,7.35)(1.5,0){5}{\vector(-1,-1){1.2}}
\multiput(11.35,5.85)(1.5,0){5}{\vector(-1,-1){1.2}}
\multiput(11.35,4.35)(1.5,0){5}{\vector(-1,-1){1.2}}
\multiput(11.35,2.85)(1.5,0){5}{\vector(-1,-1){1.2}}
\multiput(11.35,1.35)(1.5,0){5}{\vector(-1,-1){1.2}}
\end{picture}
\vspace{3ex}
  \caption{}
    \label{fig:2}
  \end{center}
\end{figure}

The following theorem summarizes our results on degenerate
$E(5,10)$-modules.

\begin{Theorem}
  \label{th:4.4}
(a)~~Each connected component in Figure~\ref{fig:2} is a complex
of $E(5,10)$-modules.

(b)~~The kernels of all morphisms in these complexes are
irreducible maximal submodules.
\end{Theorem}


\vspace{6ex}

\textbf{Authors' addresses:}
\begin{list}{}{}

\item  Department of Mathematics, MIT,
Cambridge MA 02139,
USA\\
email:~~kac@math.mit.edu

\vspace{1ex}

\item   Department~of~Mathematics, NTNU, Gl\o shaugen,
N-7491 Trondheim,
Norway \\
email:~~rudakov@math.ntnu.no

\end{list}

\end{document}